\documentclass[twocolumn,aps,prl,superscriptaddress,showpacs,floatfix,amsmath]{revtex4-2}
\usepackage{epsfig}
\usepackage{amsmath}
\usepackage{bm}
\usepackage{here}
\usepackage{float}
\usepackage{graphicx}
\usepackage[linkcolor=black,urlcolor=blue,citecolor=blue,colorlinks=true]{hyperref}
\usepackage{xcolor}

\usepackage[normalem]{ulem}

\begin{document}
	
\title{Evidence of electron correlation induced kink in Dirac bands in a non-symmorphic Kondo lattice system, CeAgSb$_2$}
	
\author{Sawani Datta$^1$, Khadiza Ali$^2$, Rahul Verma$^1$, Bahadur Singh$^1$, Saroj P. Dash$^2$, A. Thamizhavel$^1$ $\&$ Kalobaran Maiti}

\altaffiliation{Corresponding author: kbmaiti@tifr.res.in}

\affiliation{Department of Condensed Matter Physics and Materials Science, Tata Institute of Fundamental Research, Homi Bhabha Road, Colaba, Mumbai-400005, India.\\
$^2$Department of Microtechnology and Nanoscience, Chalmers University of Technology, Göteborg, 41296 Sweden.\\}
	
	
\begin{abstract}
We study the behavior of Dirac fermions in the presence of electron correlation in a nonsymmorphic Kondo lattice system, CeAgSb$_2$ employing high-resolution angle-resolved photoemission spectroscopy and first-principles calculations. Experiments reveal crossings of highly dispersive linear bands at the Brillouin zone boundary due to non-symmorphic symmetry. In addition, anisotropic Dirac cones are observed constituted by the squarenet Sb 5$p$ states forming a diamond-shaped nodal line. The Dirac bands are linear in a wide energy range with a unusually high slope and exhibit distinct Dirac point in this highly spin-orbit coupled system. Interestingly, the linearity of the bands are preserved even after the hybridization of these states with the local Ce 4$f$ states, which leads to a small reduction of slope via formation of a 'kink'. These results seed the emergence of an area of robust topological fermions even in presence of strong correlation.
\end{abstract}

\maketitle
	

Discovery of topological materials led to the emergence of a fast-growing research area due to their rich physics and wide range of applications \cite{Hasan, PtSn4}.
Topological insulators are simple bulk insulators with metallic surface states where the metallicity is  protected by time-reversal, inversion and/or other point group symmetries. Extensive research in this field revealed topological properties even in the bulk electronic structures. For example, pnictides such as Cd$_3$As$_2$ \cite{Cd3As2-1, Cd3As2-2}, Na$_3$Bi \cite{Na3Bi-1, Na3Bi-2}, NbAs \cite{NbAs}, TaAs \cite{TaAs-1, TaAs-2} and YbMnBi$_2$ \cite{YbMnBi2} exhibit signatures of Dirac and/or Weyl fermions. PbTaSe$_2$ \cite{PbTaSe2-1} and PtSn$_4$ \cite{PtSn4} show closed loop like bulk Fermi surface, defined as nodal line semimetals. The behavior of Dirac fermions in the presence of electron correlation is an open area of research with no examples reported so far. Since electron correlation enhances the effective mass of the fermions, correlated materials with robust Dirac cones may be a good test case for such studies.

Recently, systems with nonsymmorphic symmetry have drawn much attention \cite{YbMnBi2, PbTaSe2-1, Takane, Schoop, young, Schoop2}
as the symmetry operations such as screw axis, glide plane, etc. involve fractional lattice translation in addition to the point group operation. Thus, the unit cell in a non-symmorphic system becomes bigger leading to a smaller Brillouin zone (BZ), band folding and hence, band crossings at the zone boundary \cite{Klemenz}. Young and Kane showed that such nonsymmorphic symmetry-protected two-dimensional Dirac cones cannot be gapped by spin–orbit coupling (SOC), while the cases protected by other symmetries can be gapped \cite{young}. This is a nascent field with majority of materials are weakly correlated and Dirac points far away from the Fermi level, $\epsilon_F$, such as ZrSiS exhibiting Dirac point at about 0.5 eV binding energy \cite{Schoop}.

To probe the behavior of Dirac fermions in the presence of electron correlation, we studied a non-symmorphic Kondo lattice system, CeAgSb$_2$, which crystallizes in ZrCuSi$_2$-type structure (space group $P4/nmm$)
consisting of stacked [Sb1-CeSb2-Ag-Sb2Ce-Sb1] layers along [001] direction ($a$ = 4.363 \AA\ and $c$ = 10.699 \AA) \cite{Jobilong,Inada}. There are two non-equivalent Sb sites (Sb1 and Sb2); Sb2 sites are tightly packed with the Ce and Ag layers. Sb1 sites sandwiched by two Ce-layers form a squarenet structure and provide a unique platform for the interaction of the Dirac fermions of the squarenet with the strongly correlated Ce 4$f$ states.
%
%
Resistivity, $\rho(T)$ of CeAgSb$_2$ is highly anisotropic ($\rho_{[001]}/\rho_{[100]}$ = 8.3 at 300 K, and 18 at 10 K), exhibits a minima at 150 K, a Kondo-like logarithmic temperature dependence at lower temperatures, Kondo coherence below 15 K and a sharp decrease below 9.8 K \cite{Inada, Jobilong}. Magnetic measurements ($H||$[001]) show ferromagnetic order below 9.8 K with a saturation moment of 0.4 $\mu_B$/Ce. $H||$[100]-data exhibit unusual behavior; moment increases with field upto 3T and saturates to 1.2 $\mu_B$/Ce at higher field. The Kondo temperature is estimated to be 23 K. Evidently, CeAgSb$_2$ is an exotic system with the Dirac states in proximity to the strongly correlated layer. We studied the electronic structure of CeAgSb$_2$ employing high resolution angle-resolved photoemission spectroscopy (ARPES) and density functional theory (DFT) and discover exceptional properties of the Dirac fermions and their interaction with strongly correlated states.


\begin{figure*}
\centering
\includegraphics[width=1\textwidth]{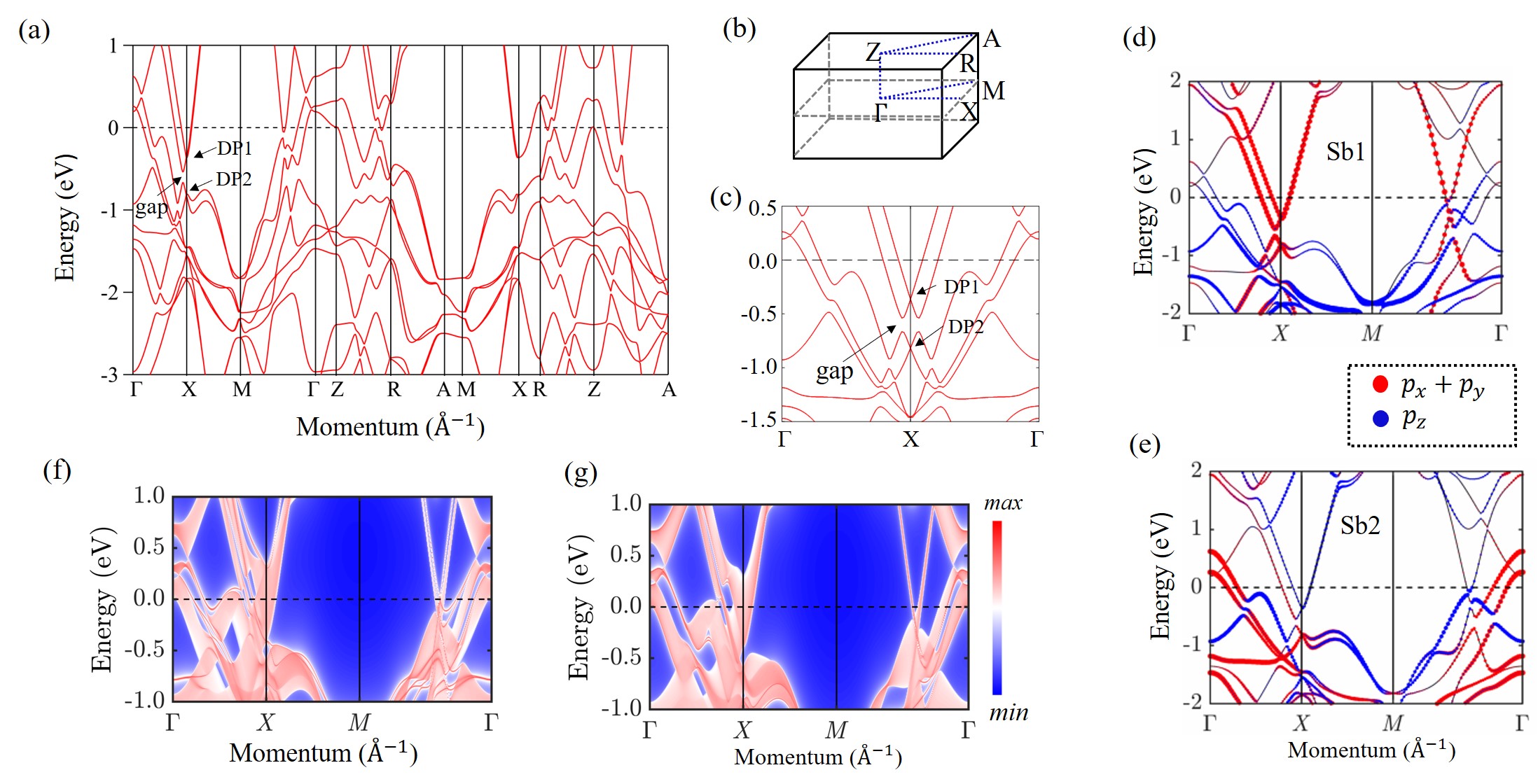}
\caption{(a) Bulk band structure including spin-orbit coupling and (b) the Brillouin zone. (c) Bands along $\Gamma X\Gamma$ in an expanded energy and momentum scale. Nonsymmorphic symmetry-protected Dirac points are marked as DP1 and DP2. The gap at the Dirac point away from $X$ is also shown. Orbital resolved bulk band structure exhibiting in-plane and out-of plane 5$p$ contributions for (d) Sb1 and (e) Sb2 sites; red and blue colors show the $p_x+p_y$ and $p_z$ orbital characters, respectively. The (001) surface projected band structure for (f) Ce and (g) Sb1 terminations.}
\label{Fig1}
\end{figure*}

High quality single crystalline samples were prepared following a solution growth method \cite{Myers}.
ARPES measurements were performed at 19 K on a cleaved surface (pressure $\sim$ 10$^{-11}$ Torr) using linear horizontal polarized light at Bloch beamline, MAX IV laboratory, Sweden using Omicron-Scienta DA30L analyzer (energy resolution, 5 meV and angle resolution, 0.1$^\circ$). Cleaved surface exposed Sb1-terminated surface as also observed earlier \cite{Sawani1} and discussed later in the text.
Electronic structure calculations were performed using projector augmented-wave (PAW) method as implemented in Vienna \textit {ab-initio} simulation package (VASP)~\cite{KohnSham1965}.
The generalized gradient approximation (GGA)~\cite{Perdew} was used to include exchange-correlation effects. The bulk and surface band structures were obtained by considering the Ce 4$f$ electrons as the core electrons. We used experimental lattice parameters with relaxed ionic positions to calculate the energy spectrums. A kinetic energy cutoff of 500 eV for the plane-wave basis set and a $\Gamma$-centered 11$\times$11$\times$9 $k$-mesh were used for the bulk BZ sampling. The surface spectrum was calculated using semi-infinite Green's function method by constructing material-specific tight-binding Hamiltonian using the VASP2WANNIER90 interface \cite{green1985}.


\begin{figure*}
\centering
\includegraphics[width=1\textwidth]{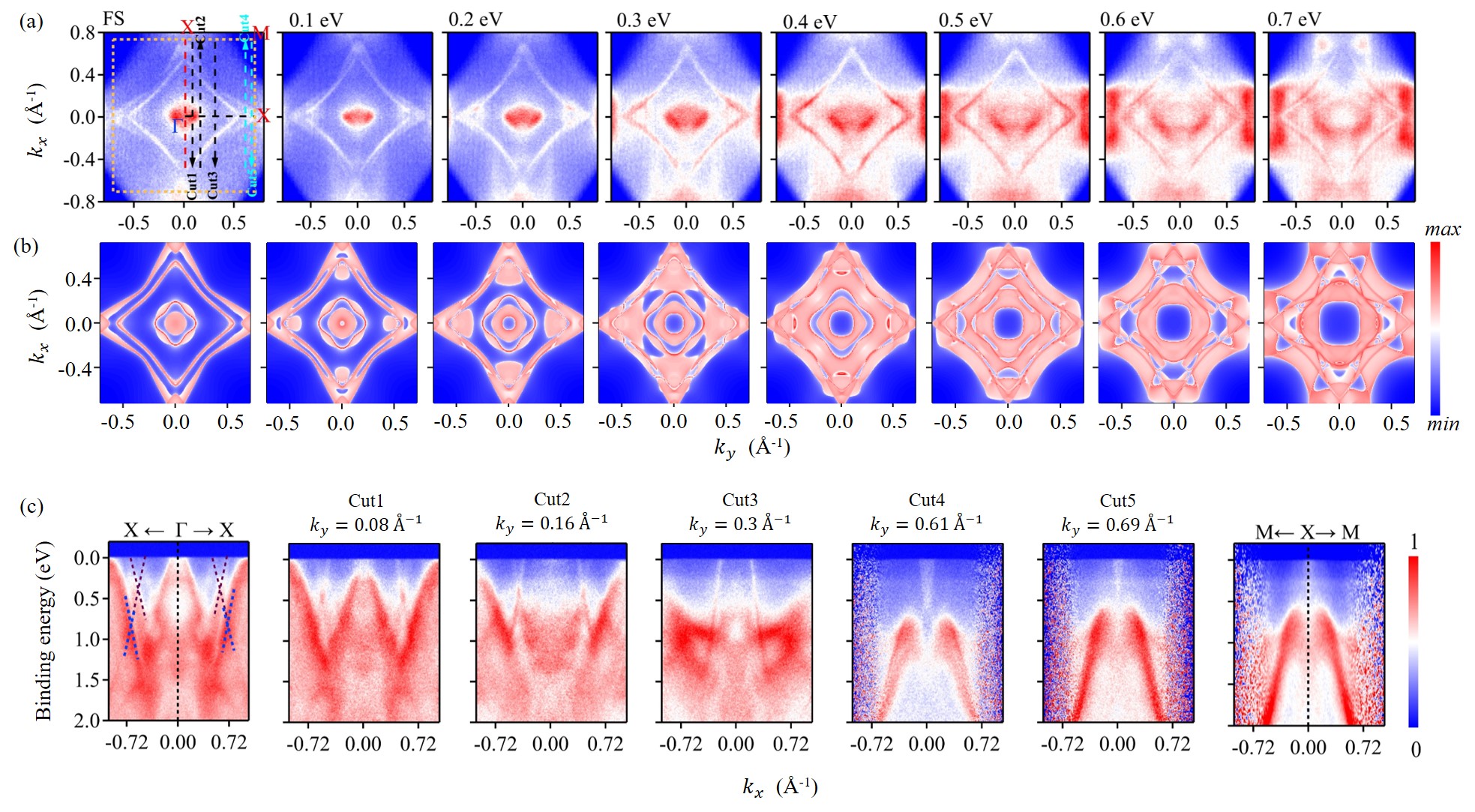}
\vspace{-4ex}
\caption{(a) Experimental and (b) calculated Fermi surface map and constant energy contours at different binding energies. The box in the left panel (FS plot) is the Brillouin zone and vertical dashed lines are the cuts at different $k_y$ values. (c) Band dispersions at the cuts shown in (a). The dashed lines in left panel of (c) are the guides to the eye showing Dirac cones.}
\label{Fig2}
\end{figure*}

From the electronic structure calculations, 
we find that Ce 4$f$ states contribute significantly near $\epsilon_F$. Sb2 states appear away from $\epsilon_F$ with negligible contributions at $\epsilon_F$. Sb1 5$p_z$ bands are essentially filled. Sb1 (5$p_x$+5$p_y$) bands have close to half-filled configuration \cite{Schoop2, Hoffmann}, which places the Dirac bands close to $\epsilon_F$ where Ce 4$f$ states also contribute making this material an excellent case to study the behavior of Dirac fermions in the presence of strong correlation. Calculated bulk band structure with spin-orbit coupling (SOC) is shown in Fig. \ref{Fig1}(a) and the high symmetry points in BZ are defined in Fig.~\ref{Fig1}(b). Several bands cross the Fermi level, have large dispersions and the dispersion along $\Gamma Z$ are relatively flat indicating quasi-2D nature of the electronic structure. Several Dirac cones are observed along $\Gamma-X$ and $\Gamma-M$ directions as well as at $X$ and $M$ points. Some of the band crossings form nodal lines as also observed in other materials such as LaAgSb$_2$\cite{Rosmus} and ZrSiS \cite{Schoop}. The band crossings at $X$ and $M$ points are protected by the glide plane and remain robust in the presence of SOC \cite{Schoop, young}. The bands along $\Gamma-X$ and $\Gamma-M$ have $C_{2v}$ symmetry and exhibit gaps at the crossings due to SOC as there is only one irreducible representation for the $C_{2v}$ group in the presence of SOC \cite{Schoop2, Schoop, ZrSiTe}. The Dirac points at $X$ are protected by nonsymmorphic symmetry and appear close to $\epsilon_F$ as envisaged for this system. For clarity, we plot the band structure along $\Gamma-X-\Gamma$ in an expanded energy and momentum scales in Fig.~\ref{Fig1}(c). The Dirac points at $X$ marked by DP1 and DP2, and the energy gap away from $X$ are clearly visible in the figure. Similar Dirac points are also observed at $R$ with a slightly larger energy separation. The Dirac crossings have predominantly Sb1 5$p_x$+5$p_y$ orbital contribution, as visible in the orbital character plots of the bulk bands shown in Figs.~\ref{Fig1} (d) and (e). The Sb2 5$p_x$+5$p_y$ orbital will form two hole pockets around the $\Gamma$ point. The 5$p_z$ bands of both Sbs are lying below the Fermi level.

\begin{figure}
\centering
\includegraphics[width=0.5\textwidth]{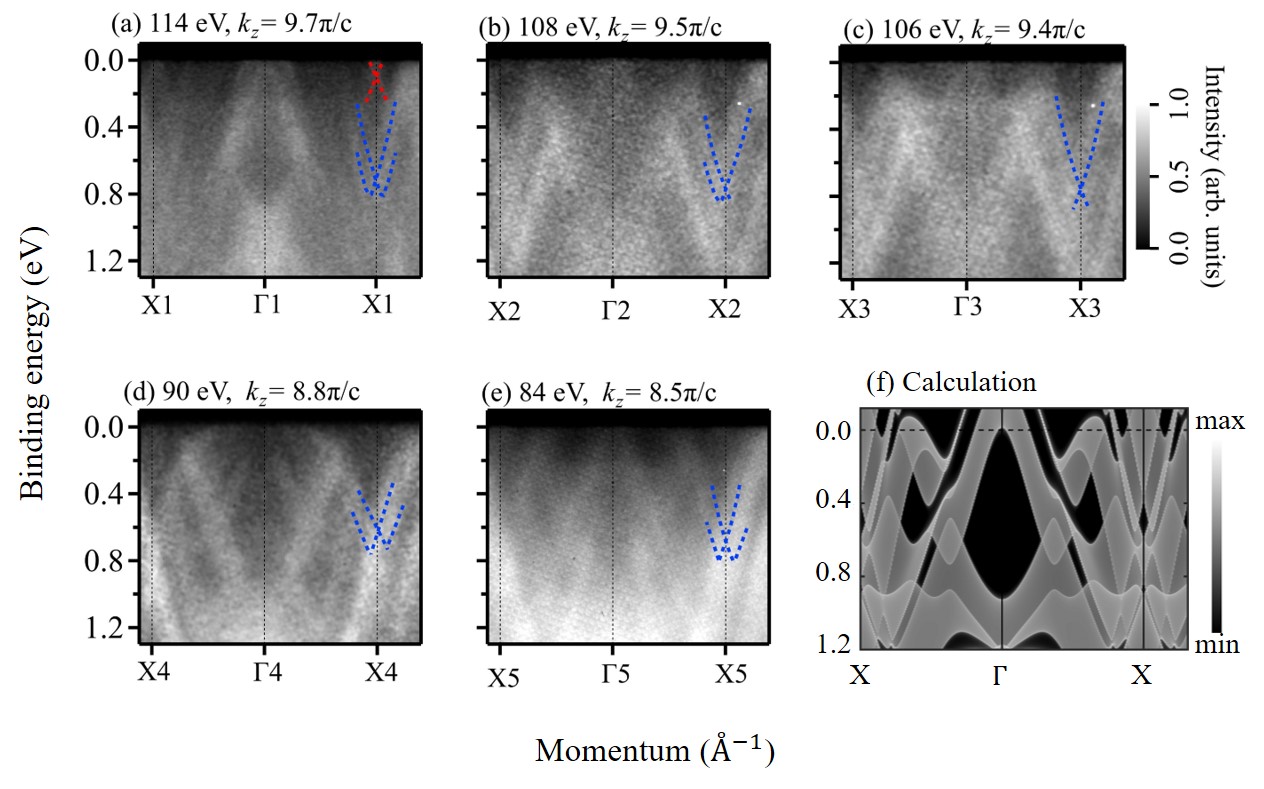}
\vspace{-4ex}
\caption{(a)-(e) Experimental band structure along $X\Gamma X$ at different $k_z$ values; dashed lines are guide to eye. (f) Calculated bulk bands along $X\Gamma X$ projected on the (001) surface.}
\label{Fig3}
\end{figure}

In Figs.~\ref{Fig1}(f) and (g), we show the (001) surface band structure with Ce and Sb1 terminations, respectively. Projection of the bulk bands on the surface Brillouin zone exhibit highly dispersive broad bands. The surface bands, shown by dark red color, exhibit dependence on the termination layer. While the surface bands near $\Gamma$-point is somewhat similar in both the cases, Ce-terminated case exhibits more resolved surface bands along $\Gamma-X$ and $\Gamma-M$ lines as evident in Fig.~\ref{Fig1}(f). The Sb-terminated case shown in Fig.~\ref{Fig1}(g) exhibits less complex scenario with significantly less number of distinct surface bands. There is an electron pocket along $\Gamma-M$; the corresponding bands show a Dirac cone like structure below $\epsilon_F$. In the Sb-terminated case, the size of the Fermi pocket is larger. These results suggest that the properties of Ce 5$d$ states have more profound effect due to surface termination with significantly different surface and bulk electronic structures in the Ce-terminated case, while Sb1 5$p$ states are relatively less affected by the surface termination due to their effectively two-dimensional structure.

\begin{figure}
\centering
\includegraphics[width=0.5\textwidth]{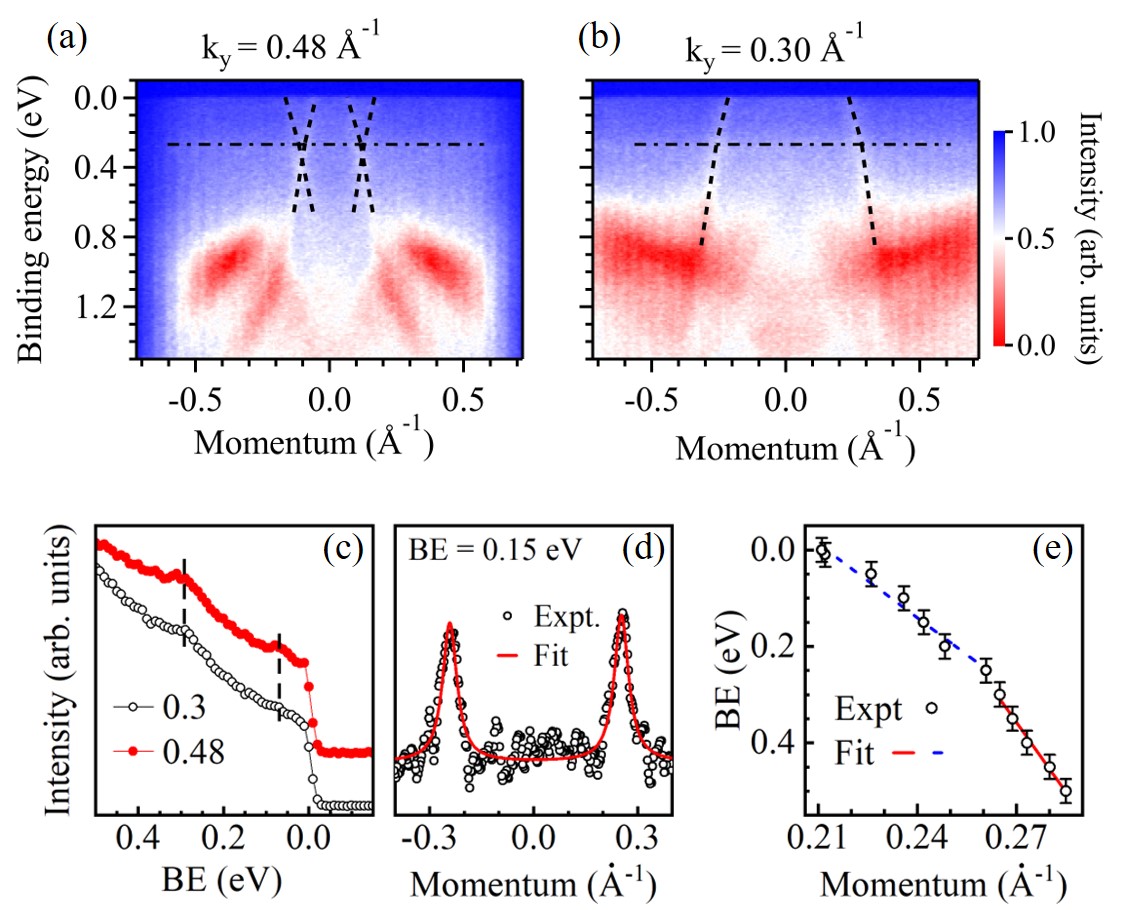}
\caption{Band structure parallel to $X-\Gamma-X$ at $k_y$ values of (a) 0.48 \AA$^{-1}$ and (b) 0.30 \AA$^{-1}$; dashed lines are guide to eyes. Horizontal dot-dashed line is Ce 4$f$-band. (c) EDC for the $k$-integration range -0.7 to -0.3 \AA$^{-1}$ showing Ce 4$f$-signal. (d) MDC at 0.15 eV (symbols) and fit (line). (e) MDC ($k_y$ = 0.30 \AA$^{-1}$) peak positions (points). The lines are linear fits.}
\label{Fig4}
\end{figure}

ARPES measurements carried out on freshly cleaved sample exhibit band structure closely resembling the case of the Sb1 termination; we did not observe additional surface bands typical of the Ce-terminated case consistent with earlier observations \cite{Sawani1}. In Fig.~\ref{Fig2}(a), we plot the Fermi surface and the constant energy contours measured at different binding energies (BEs) using 52 eV photon energy. Corresponding calculated results are shown in Fig.~\ref{Fig2}(b). In the Fermi surface plot, we observe two intense diamond-shaped large Fermi pockets due to the surface states. These pockets marge at 0.4 eV and then becomes distinct again at higher binding energies. Other than the diamond sheets, two Fermi pockets around the $\Gamma$ point are also visible. A small pocket around $X$ appears just below the Fermi surface. These pockets gradually grow in size at higher binding energies.

To probe the property of the single-sheet-diamond-shaped pocket at 0.4 eV, we study the band dispersions collected along the cuts at different $k_y$-values shown in the left panel of Fig.~\ref{Fig2}(a). The band structure along $X\Gamma X$ exhibits linearly dispersed bands (marked by dashed line) crossing each other at 0.4 eV forming a Dirac cone. These Dirac cones move closer to each other with the increase in $k_y$-values and merge at $X$ point as evident in the $M-X-M$ data consistent with the calculated results. These results demonstrate the diamond-shaped nodal line in CeAgSb$_2$. Interestingly, such $C_{2v}$ symmetry-protected Dirac cones are expected to show an energy gap at the Dirac point in this strongly spin-orbit coupled system. Thus, the presence of distinct crossings in the experimental data is a deviation from such theory may be due to the experimental (resolution and lifetime) broadenings and/or interaction with the surface states leading to a more complex scenario demanding further study. The nonsymmorphic symmetry-protected crossing, DP2 appears at 0.8 eV. These Dirac bands exhibit linearity over a wide energy range (almost 1.5 eV), significantly larger than the observations in other materials \cite{Schoop}. The bands close to $\epsilon_F$ show deviations from DFT results indicating a role of strong correlation due to the hybridization with the Ce 4$f$ states.

In Fig. \ref{Fig3}, we show the nonsymmorphic symmetry-protected Dirac bands probed using different photon energies for the $\Gamma_i$(0,0,$k_z$) - $X_i$($\pi/a$,0,$k_z$) vectors. The Dirac point at 0.8 eV at $X$ remains almost the same at all the photon energies indicating it's insensitivity to $k_z$ variation. This matches well with the surface projected bulk band structure shown in Fig.~\ref{Fig3}(f). The absence of $k_z$ dependence manifests the effective 2D nature of the electronic structure as also reflected in the bulk properties measurements. In Fig.~\ref{Fig3}(a), we observe an additional Dirac cone, DP1 close to $\epsilon_F$. The energy separation between DP1 and DP2 is about 0.7 eV, which is somewhat larger than the separation observed at $X$ in the calculated bulk data shown in Fig.~\ref{Fig1}(c) and smaller than the value at $R$. Since the $k_z$ corresponding to this photon energy lies between 0 to $\pi/c$, such separation is consistent with the expected behavior.

Now, we study the role of electron correlation on the properties of nodal line Dirac bands in proximity to the Ce-layer. The Dirac bands for $k_y$ = 0.48 \AA$^{-1}$ along $X\Gamma X$-direction shown in Fig.~\ref{Fig4}(a) exhibit highly dispersive behavior and band crossing at 0.4 eV as discussed above. The Dirac bands maintain linearity in a large energy window with a significantly large slope. Interestingly, the slope reduces slightly in the near $\epsilon_F$-regime preserving the linearity of the bands. This behavior is also evident in the data for $k_y$ = 0.3 \AA$^{-1}$ shown in Fig.~\ref{Fig4}(b) where one branch of the Dirac cone becomes dark due to the matrix element effect \cite{Han}. To investigate this scenario further, we derived the peak positions of the momentum distribution curves (MDCs) of the data in Fig.~\ref{Fig4}(b) (a typical fitting is shown in Fig.~\ref{Fig4}(d)) and plot in Fig.~\ref{Fig4}(e). The results exhibit a clear change in slope at about 0.28 eV where Ce 4$f$ band appears as shown by the horizontal dot-dashed line in Figs.~\ref{Fig4}(a) and (b) and the energy distribution curves (EDCs) in Fig. \ref{Fig4}(c) obtained by integrating the data in the $k$-range -0.3 to -0.7 \AA$^{-1}$. The slope, $\hbar v_F$ near $\epsilon_F$ is found to be 5.1 eV.\AA, which is larger than ZrSiS (= 4.3 eV.\AA) \cite{Schoop} and smaller than that in graphene (= 6.7 eV.\AA). At higher binding energies (BE $>$ 0.28 eV), the slope of the band increases significantly to about 9.5 eV.\AA, indicating an exceptionally high mobility of these topological fermions, much higher than graphene. These results reveal highly relativistic nature of the Sb 5$p$ electrons within the squarenet structure and formation of a kink due to electron correlation, which is outstanding. The decrease in the slope near $\epsilon_F$ is attributed to hybridization between the Sb 5$p$ states with the strongly correlated Ce 4$f$ states \cite{Swapnil2}. Most interestingly, the relativistic behavior of the fermions did not change even after hybridization with the strongly correlated states. Usually kinks are observed due to electron-phonon interaction which is often used as an evidence of role of electron-phonon coupling in the superconductivity of the material. This discovery of the Dirac band bending forming a 'kink' due to strong correlation and survival of the relativistic properties is unique and provides an unique example of the interplay of electron correlations and topological protection signalling emergence of a new direction of thoughts in exotic materials.


In summary, the band structure of CeAgSb$_2$ reveals distinct nonsymmorphic symmetry-protected Dirac cones at high symmetry points. A diamond-shaped nodal line due to squarenet structure is observed in this strong spin-orbit coupled system. Linearly dispersed bands exhibit an unusually high electron velocity, much higher than graphene and form two-sheet-diamond-shaped Fermi pockets. Interestingly, the velocity reduces in proximity of the strongly correlated Ce 4$f$ bands forming a 'kink' manifesting a unique scenario of the behavior of topological fermions in the presence of strong correlation, which is expected to open up an exciting area of research in topological quantum materials.

\textbf{Acknowledgements:} Authors thank the Dept. of Atomic Energy, Govt. of India (Project no. RTI4003, DAE OM no. 1303/2/2019/R\&D-II/DAE/2079) for financial support and MAX IV Lab. for the time at Bloch beamline (Proposal 20220940). Research at MAX IV is supported by the Swedish Research Council under contract 2018-07152, the Swedish Govt. Agency for Innovation Systems under contract 2018-04969, and Formas under contract 2019-02496.

\end{document}